\documentclass[11pt,singlecolumn]{article}
\usepackage[top=0.8in, bottom=0.8in, left=0.8in, right=0.8in]{geometry}
\usepackage{graphicx}
\usepackage{amssymb}
\usepackage{amsmath}
\usepackage{colortbl}
\usepackage{color}
\usepackage[small,bf]{caption}
\usepackage{indentfirst}
\usepackage{url}
\usepackage{subfigure}
\usepackage{multirow}
\usepackage{threeparttable}
\usepackage{tikz}
\usepackage{bbding}
\usepackage[all]{xy}
\usepackage{algorithmic}
\usepackage{array}
\usepackage{multirow}
\usepackage{flushend}
\usepackage{paralist}
\usepackage{xspace}
\usepackage{mathptmx}
\usepackage{anyfontsize}
\usepackage{t1enc}
\definecolor{darkgreen}{RGB}{0,0,150}
\usepackage[bookmarks=false, colorlinks=true, plainpages=false,  linkcolor=darkgreen,   citecolor=darkgreen, urlcolor=darkgreen, filecolor=darkgreen]{hyperref}
\usepackage{breakurl}
\usepackage{paralist}
\usepackage[verbose]{wrapfig}

\usepackage[compact]{titlesec}
\titlespacing*{\section}{0pt}{*3}{3.5pt}
\titlespacing{\subsection}{0pt}{*3.5}{3.5pt}
\titlespacing{\subsubsection}{0pt}{*1.5}{0pt}

\usepackage{indentfirst}

\usepackage[hang,flushmargin,stable]{footmisc}

\date{}

\begin{document}

\title{\bf Privacy-Related Consequences of Turkish Citizen Database Leak}

\author{Erin Avllazagaj$^1\;\;\;\;$ Erman Ayday$^2\;\;\;\;$ A. Ercument Cicek$^3$\\[1.5ex]
$^1$ Bilkent University, erin.avllazagaj@ug.bilkent.edu.tr\\[0.2ex]
$^2$ Bilkent University, erman@cs.bilkent.edu.tr\\[0.2ex]
$^3$ Bilkent University and Carnegie Mellon University, cicek@cs.bilkent.edu.tr
}

%
%

\maketitle

\begin{abstract}
Personal data is collected and stored more than ever by the governments and companies in the digital age. Even though the data is only released after anonymization, deanonymization is possible by joining different datasets. This puts the privacy of individuals in jeopardy. Furthermore, data leaks can unveil personal identifiers of individuals when security is breached. Processing the leaked dataset can provide even more information than what is visible to naked eye. In this work, we report the results of our analyses on the recent "Turkish citizen database leak", which revealed the national identifier numbers of close to fifty million voters, along with personal information such as date of birth, birth place, and full address. We show that with automated processing of the data, one can uniquely identify (i) mother's maiden name of individuals and (ii) landline numbers, for a significant portion of people. This is a serious privacy and security threat because (i) identity theft risk is now higher, and (ii) scammers are able to access more information about individuals. The only and utmost goal of this work is to point out to the security risks and suggest stricter  measures to related companies and agencies to protect the security and privacy of individuals.
\end{abstract}

\section{Introduction}

Personal data is crucial for companies, governments, and service providers. Data collected by the service providers about the individuals increases the utility of the services they provide. As a result, today, an adult's records exists in more than 700 databases around the World~\cite{databases}.

Companies or governments either (publicly) share personal data after anonymization (e.g., for research purposes) or they keep it locally for their own records. In today's interconnected world, even if a dataset is shared after anonymization (i.e., removal of personal identifiers from the data such as the name or telephone number), in many cases, it is possible to deanonymize the dataset participants by using auxiliary information (that is publicly available).

Data leaks are also major privacy threats for individuals. Leaked datasets alone may include considerable amount of information about the individuals (especially, if it is not anonymized). On top of this, an attacker can infer sensitive information about the individuals in the leaked dataset (that cannot be directly observed from the dataset). Even worse, an attacker can use auxiliary information (e.g., online social networks or publicly available datasets) to infer more about the individuals that are in the leaked dataset (and about their family members, friends, etc.).

In this work, we report our findings about the privacy-related threats of the ``Turkish citizen database leak''.  Personal records of 49,611,709 Turkish citizens have publicly become available a few months ago~\cite{turkish_leak}. Experts confirmed this leak as one of the biggest public leaks of personal data ever seen, as it puts almost two-thirds of the country's population at risk of fraud and identity theft. The leak actually happened before 2010, but it recently became publicly available. The leaked dataset includes voter registration data of the citizens for the 2009 local elections and reveals sensitive information about the citizens including their personal details and national identifiers (equivalent of the SSN in the US). Potential attacks by directly using such information of the citizens include (i) generating fake identity cards, (ii) loaning money from a bank, (iii) getting phone lines, (iv) founding a company, and (v) being a guarantor.

We mainly study two important aspects of this leaked database: (i) the uniqueness of the citizens (inspired by the well-known study of Latanya Sweeney on the US census data~\cite{Sweeney2000}) to show the comparative threat of disclosing various demographic data, and (ii) potential further inference about unrevealed sensitive data of the individuals. For the latter, we focus on determining ``mother's maiden name'' and ``personal landline number'' of each individual in the dataset.

Mother's maiden name has been a security question since 1882~\cite{maiden_name}. It is commonly used as a remote authentication question by many service providers. Especially financial institutions use this type of  ``knowledge-based authentication'' very commonly because the answer does not change over time, the question has an answer for all the users, users can easily remember the answer, and the institutions think that it is safe (i.e., hard to guess by an attacker). However, in today's internet era, people share vast amount of information on numerous online services, and hence much of this knowledge-based information becomes available to the attackers via online social networks.

If an attacker gets to learn a victim's mother's maiden name, for example, he can convince the customer service representative of a bank to wire funds from the victim's bank account to his account, a simple consequence that recently happened in the US~\cite{maiden_name2}. Note that even if a customer service representative gets suspicious, the attacker can always call again to try a different one. Similarly, if the attacker is unsure about the sensitive information (e.g., if the attacker can narrow down the potential candidates for the mother's maiden name to 2), he can also try more than once.

Learning one's mother's maiden name through social networks, even though is possible, is a targeted attack. In other words,  it is not scalable (the attacker needs to spend a lot of effort to learn this information for thousands or millions of individuals). However, as we will show in this work, with this leaked dataset, it is possible to infer the mother's maiden name of thousands of individuals.\footnote{Note that this information is not directly available in the dataset, but we use an inference technique to infer this information for each individual.}

Similarly, learning the personal landline numbers of individuals (whose personal information is available in the leaked dataset) also poses a significant threat, especially considering the widespread phone scams. In such phone scams, scammers use the terms ``you won a prize'', ``you have some dept for your health insurance'', ``we are calling from the law enforcement, it seems that some criminal activity has occurred in your bank account'', or ``there is a court order against you'' on the phone and they request money or credit card information from the individuals. Of course, they, especially if they claim that they are calling from the law enforcement or the court, provide more information about the victim to make their scenario stronger. Using sensitive information that is directly available from the leaked dataset (e.g., national identifier or full address) or inferred from the leaked dataset (e.g., mother's maiden name, as we will show in this work) throughout the call makes the victim less suspicious about the attacker, and hence the victim becomes more prone to giving what the attacker wants. In this work, we will also show how the scammers can learn the landline numbers of the individuals in the database systematically by using other publicly available sources on the Internet.

The rest of this paper is organized as follows. In the next section, we summarize the known database leaks and inference attacks. In Section~\ref{sec:format}, we describe the format of the leaked dataset. In Section~\ref{sec:threat}, we discuss the threat model. In Section~\ref{sec:uniqueness}, we present our findings about the uniqueness of the citizens based on different demographics. In Section~\ref{sec:attack}, we describe two inference attacks by using the leaked dataset. In Section~\ref{sec:discussion}, we discuss the potential countermeasures and future work. Finally, in Section~\ref{sec:conclusion}, we conclude the paper.

\section{Related Work}\label{sec:related_work}

There have been many data beaches (intentional or unintentional release of secure information) from well-known companies in the recent years. In 2014, a collection of about 500 photos of several celebrities have been leaked from iCloud service of Apple~\cite{icloud}. The main reasons of this leak were (i) weak passwords of the users, (ii) easy-to-guess security questions, and (iii) a bug in Apple's photo backup service (so that hackers were able to force their way into celebrities' photo collections by repeatedly guessing passwords). Similarly, a recent data breach exposed user account information of around 38 million Adobe users~\cite{adobe}. As a result of a well-known data breach from Target, hackers obtained credit card information of around 40 million Target customers~\cite{target}. Target had to reach a \$39 million settlement with several US banks due to this breach~\cite{target_settlement}. Furthermore, according to Office of Civil Rights (OCR), there have been 253 health data breaches that affected 500 individuals or more with a combined loss of over 112 million health records in just 2015~\cite{health}. Thus, Turkish citizen data leak is not the only data leak in the World, but it is one of the most serious leaks to date.

Hackers or attackers have many techniques to obtain unauthorized information from the leaked or publicly shared (anonymized) data. The most well-known technique that can be used to learn more about individuals is profile matching (or deanonymization). Studies show that in today's digital world, anonymization is not an effective way of protecting sensitive data. For example, Latanya Sweeney showed that it is possible to de-anonymize individuals by using publicly available anonymized health records and other auxiliary information that can be publicly accessed on the Internet~\cite{Sweeney2002}. It has been shown that anonymization is also an ineffective technique for sharing genomic data~\cite{Gitschier2009,Gymrec_Science}. For instance, genomic variants on the Y chromosome are correlated with the last name (for males). This last name can be inferred using public genealogy databases. With further effort (e.g., using voter registration forms) the complete identity of the individual can also be revealed~\cite{Gymrec_Science}. Also, unique features in patient-location visit patterns in a distributed health care environment can be used to link the genomic data to the identity of the individuals in publicly available records~\cite{Malin_boi_2004}.

Furthermore, Narayanan and Shmatikov proposed an efficient graph-based deanonymization framework based on seed and extend technique~\cite{narayanan2009anonymizing}. Ji et. al proposed another structural data deanonymization framework that does not require the seed information~\cite{ji2014ccs}. Ji et. al also proposed an evaluation system for graph anonymization and deanonymization named SecGraph~\cite{ji2015usenix}. Note that in this work, we do not need a deanonymization technique, as the leaked data already includes the identifiers of the individuals. But, this dataset can be used as auxiliary data to deanonymize the records of individuals in other datasets.

\section{Database Format}\label{sec:format}

As mentioned, the leaked database includes the records of nearly 50 million citizens (approximately 7GB of data). For each citizen, the database includes the fields below:

\begin{itemize}
  \item National Identifier
  \item First Name
  \item Last Name
  \item Mother's First Name
  \item Father's First Name
  \item Gender
  \item City of Birth
  \item Date of Birth
  \item ID Registration City and District
  \item Full Address (door number, street, neighborhood, district, and city)
\end{itemize}

Note that all individuals in the dataset are older than 18 as the dataset includes voter registration data. In the following sections, we will show how one can exploit the data in these fields to infer more information about the citizens in the database.

\section{Threat Model}\label{sec:threat}

In this work, we consider a passive attacker. That is, the attacker can only access publicly available information. This includes the leaked dataset itself, and other public resources that are available on the Internet. We do not consider any other background information for the attacker (e.g., the attacker personally knowing some individuals from the dataset, or the attacker buying information about some individuals that are in the dataset).  We also assume that the attacker is computationally bounded. That is, the attacker can only run polynomial-time algorithms to infer data about the individuals.

\section{Uniqueness of Turkish Citizens}\label{sec:uniqueness}

Latanya Sweeney showed that 87$\%$ of the US population can be uniquely identified by gender, ZIP code, and full date of birth based on 1990 US census data~\cite{Sweeney2000}. Later, using the more recent 2000 census data, another study showed that these 3 pseudo-identifiers can only uniquely identify 63$\%$ of the US population~\cite{Golle2006}. Such studies are in particularly important for researchers and data collectors that need estimates of the threat of disclosing anonymized datasets along with pseudo-identifiers (e.g., demographic data).

In this part, we do a similar study for the Turkish citizens using the leaked dataset and compare our results with the previous work. The result of this study shows the fraction of Turkish citizens that can be uniquely identified by using combinations of different pseudo-identifiers, and hence it gives some insight about how to release anonymized datasets of Turkish citizens.

For this study, we use the following pseudo-identifiers (and their generalizations) that are available in the dataset: (i) gender ($G$), (ii) full date of birth ($D$), (iii) city of birth ($C$), and (i) full address ($A$). The pseudo-identifier set we use in this study can be represented as $\mathbf{P}=\{G_i^j,D_i^j,C_i^j,A_i^j\}$, where $i$ represents the generalization level of the corresponding pseudo-identifier, and $j$ denotes whether the corresponding pseudo-identifier is used in a given experiment ($j\in\{0,1\}$, 1 means the pseudo-identifier is used, and 0 means otherwise). For a given pseudo-identifier, $i=0$ represents the ground level, at which no generalization made. Next, we briefly discuss our generalization assumptions.

Full address includes door number, street, neighborhood, district, and city when $i=0$. One level generalization ($i=1$) only includes street, neighborhood, district, and city. Two level generalization ($i=2$) only includes neighborhood, district, and city. Three level generalization ($i=3$) only includes district, and city. Four level generalization ($i=4$) only includes the city. Full date of birth includes day, month, and year when $i=0$. One level generalization ($i=1$) only includes the month and the year. Two level generalization ($i=2$) only includes the year. We do not generalize gender and the city of birth ($i$ values is always 0 for $G$ and $C$). In the following, we report some of our notable findings.
\begin{itemize}
  \item When $\mathbf{P}=\{G_0^1,D_0^1,C_0^1,A_0^1\}$, we could uniquely identify $93.6\%$ of the Turkish citizens.
  \item When $\mathbf{P}=\{G_0^1,D_0^1,C_0^0,A_0^1\}$, we could uniquely identify $93.4\%$ of the Turkish citizens.
  \item When $\mathbf{P}=\{G_0^1,D_0^1,C_0^0,A_2^1\}$ (the most similar case to the previous studies on US census data), we could uniquely identify $67.9\%$ of the Turkish citizens.
\end{itemize}

The rest of the results for (i) $\mathbf{P}=\{G_0^1,D_0^1,C_0^0,A_i^1\}$ for $i=\{0,1,2,3,4\}$ and (ii) $\mathbf{P}=\{G_0^1,D_i^1,C_0^0,A_0^1\}$ for $i=\{0,1,2\}$ are also shown in Table~\ref{table:uniqueness}. From these results, we conclude that both the address and the date of birth should be generalized for at least 2 levels when disclosing anonymized datasets (that include these pseudo-identifiers).
\begin{table}[ht]
\centering
\includegraphics[scale=0.75]{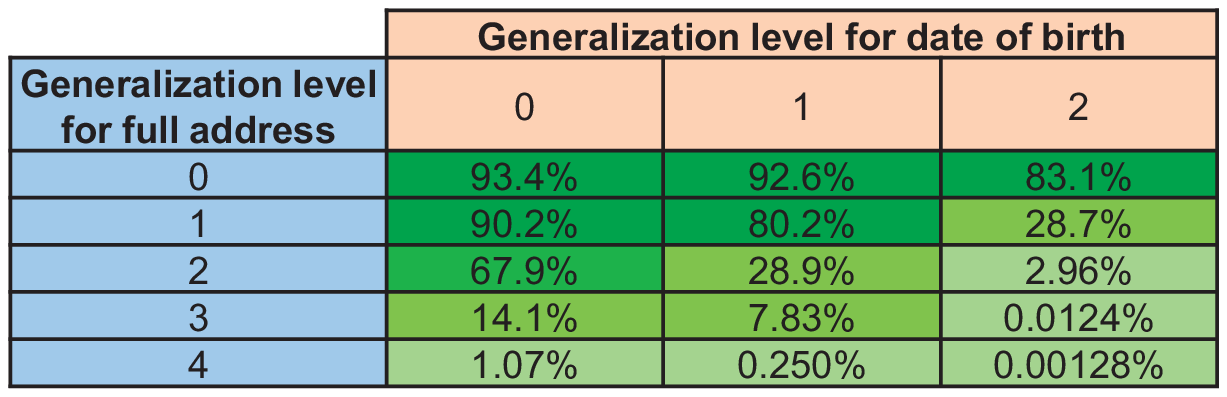}
\caption{Uniqueness of the citizens with respect to the generalization level for date of birth and full address.}
\label{table:uniqueness}
\end{table}

\section{Attacks and Results}\label{sec:attack}

In this section, we show two potential attacks via further inference from the revealed dataset about unrevealed sensitive data of the individuals.

\subsection{Inferring Mother's Maiden Name}\label{sec:maiden}

As discussed, one's mother's maiden name is a very popular knowledge-based authentication technique that is used by several service providers (especially the financial institutions). This information is not directly available in the leaked dataset that we consider. In this section, we show that using the leaked dataset (or using similar publicly available datasets), it is possible to infer this information for individuals without using any other auxiliary information.

The attacker does not need to uniquely determine the mother's maiden name of a victim via this attack. As we mentioned, if the attacker is unsure about the sensitive information, he can easily try multiple times (e.g., through different customer service representatives). Thus, in this attack, the ultimate goal of the attacker is to reduce the size of the anonymity set as much as possible for a victim, and hence we evaluate the power of this attack by analyzing the size of the anonymity set. We denote the victim as $v$ (i.e., individual whose mother's maiden name is inferred by the algorithm) and the full address of the individual $i$ as $\mathrm{Addr_i}$. The attacker runs Algorithm~1 (which either calls Algorithm~2 or~3 depending on the type of the victim) for each victim. As a result, the attacker gets the result set (i.e., anonymity set) for each victim.

\noindent\textbf{Algorithm~1 (finding the mother's maiden name of an individual)}\vspace{-6pt}
\begin{enumerate}
  \item Initialize the result set $G$ and intermediate result set $R$ as empty
  \item Get the victim $v$
  \item If the victim is female, infer the marital status using the address $\mathrm{Addr_v}$\footnote{If living alone or living with her parents, we assume the victim is single; if living with her husband and no parents, we assume the victim is married.}
  \item If the victim is male or single female go to Algorithm~2
  \item If the victim is a married female go to Algorithm~3
  \item Return the result set $G$
\end{enumerate}

\noindent\textbf{Algorithm~2 (finding the mother of a male or single female citizen)}\\
Assumption~1: Male or single female victim has the same last name as his/her father\\
Assumption~2: The mother is at least $H$ years older than the victim\footnote{We run the algorithm for three different values of $H$ (15, 25, and 35).}
\begin{enumerate}
  \item Input: victim $v$ and set $R$
  \item Initialize sets $F$ and $M$ as empty
  \item Set target person $t$ to $v$
  \item Get the last name of the target: $\mathrm{last\_name}$
  \item Get the first name of target's father: $\mathrm{father\_first}$
  \item Get the first name of target's mother: $\mathrm{mother\_first}$
  \item Get the address of the target: $\mathrm{Addr_t}$\\
  \noindent//If father and mother are married
  \item Find all males with $\mathrm{father\_first}$  \& $\mathrm{last\_name}$  and construct $F$
  \item Find all females with $\mathrm{mother\_first}$  \& $\mathrm{last\_name}$  and construct $M$
  \item Compare $\mathrm{Addr_t}$ with the addresses of individuals in $F$ and $M$
  \item If there is an individual in $M$ with the same address as $\mathrm{Addr_t}$, add her to $R$ (target living with his/her parents in the same house)
  \item Else if there is an individual in $M$ with a different address than $\mathrm{Addr_t}$, but with the same address with an individual in $F$, add her into $R$ (target living away from his/her parents, parents are living together)
  \item If $R$ is non-empty, go to Algorithm~4 using the females in $R$ and victim $v$ as input\\
  \noindent// If father and mother are divorced or not married
  \item If $R$ is still an empty set
  \item Find all males with $\mathrm{father\_first}$  \& $\mathrm{last\_name}$  and construct $F$
  \item Find all females with $\mathrm{mother\_first}$  \& $\mathrm{any\_last\_name}$  and construct $M$
  \item Compare $\mathrm{Addr_t}$ with the addresses of individuals in $F$ and $M$
  \item If there is an individual in $M$ with a different address than $\mathrm{Addr_t}$, but with the same address with an individual in $F$, add her into $R$ (unmarried couples with children)
  \item Else if there is an individual in $M$ with the same address as $\mathrm{Addr_t}$, add her to $R$ (divorced parents, child living with the mother)
  \item If $R$ is non-empty
  \item Set $G=R$ (females in $R$ already have their maiden names)
  \item If $R$ is still empty return unsuccessful search
\end{enumerate}

\noindent\textbf{Algorithm~3 (finding the mother of a married female citizen)}\\
Assumption~1: Female victim has the last name of her husband; not the last name as her father\\
Assumption~2: The mother is at least $H$ years older than the victim\footnote{We run the algorithm for three different values of $H$ (15, 25, and 35).}
\begin{enumerate}
  \item Input: victim $v$ and set $R$
  \item Initialize sets $F$ and $M$ as empty
  \item Set target person $t$ to $v$
  \item Get the first name of target's father: $\mathrm{father\_first}$
  \item Get the first name of target's mother: $\mathrm{mother\_first}$
  \item Get the address of the target: $\mathrm{Addr_t}$\\
  \noindent//If father and mother are married
  \item Find all males with $\mathrm{father\_first}$  \& $\mathrm{any\_last\_name}$  and construct $F$
  \item Find all females with $\mathrm{mother\_first}$  \& $\mathrm{any\_last\_name}$  and construct $M$
  \item Compare $\mathrm{Addr_t}$ with the addresses of individuals in $F$ and $M$
  \item If there is an individual in $M$ with the same address as $\mathrm{Addr_t}$, add her to $R$ (target living with her mother in the same house)
  \item Else if there is an individual in $F$ with the same address as $\mathrm{Addr_t}$, add the corresponding individual from M to $R$ (target living with her father in the same house)
  \item Else if there is an individual in $M$ with a different address than $\mathrm{Addr_t}$, but with the same address with an individual in $F$, add her into $R$ (target living away from her parents, parents are living together)
  \item If $R$ is non-empty, go to Algorithm~4 using the females in $R$ and victim $v$ as input
  \item If $R$ is empty, return unsuccessful search
\end{enumerate}

\noindent\textbf{Algorithm~4 (finding the father of victim's mother)}\\
Assumption~1: Victim's mother has the last name of her husband; not the last name as her father\\
Assumption~2: The father is at least $H$ years older than the daughter\footnote{We run the algorithm for three different values of $H$ (15, 25, and 35).}
\begin{enumerate}
  \item Input: victim $v$ and set $R$
  \item Initialize set $R'$ as empty
  \item Initialize sets $F$ and $M$ as empty
  \item For each individual $i$ in set $R$, set the target person $t$ to $R(i)$
  \item Get the first name of target's father: $\mathrm{father\_first}$
  \item Get the first name of target's mother: $\mathrm{mother\_first}$
  \item Get the address of the target: $\mathrm{Addr_t}$\\
  \noindent//If father and mother are married
  \item Find all males with $\mathrm{father\_first}$  \& $\mathrm{any\_last\_name}$  and construct $F$
  \item Find all females with $\mathrm{mother\_first}$  \& $\mathrm{any\_last\_name}$  and construct $M$
  \item Compare $\mathrm{Addr_t}$ with the addresses of individuals in $F$ and $M$
  \item If there is an individual in $M$ with the same address as $\mathrm{Addr_t}$, add the corresponding individual from $F$ to $R'$ (target living with her mother in the same house)
  \item Else if there is an individual in $F$ with the same address as $\mathrm{Addr_t}$, add him to $R'$ (target living with her father in the same house)
  \item Else if there is an individual in $F$ with a different address than $\mathrm{Addr_t}$, but with the same address with an individual in $M$, add him into $R'$ (target living away from her parents, parents are living together)
  \item $G = G \cup R'$ and go to 4
  \item If $R'$ is still an empty set return unsuccessful search
\end{enumerate}

For the evaluation, we randomly sampled around 20000 individuals from the dataset and set them as victims. Among the sampled individuals, we made sure that the gender distribution and the age distribution are uniform. We ran Algorithm~1 on all the sampled individuals and constructed the anonymity set for each victim.

\begin{figure}[ht]
\centering
\includegraphics[scale=0.75]{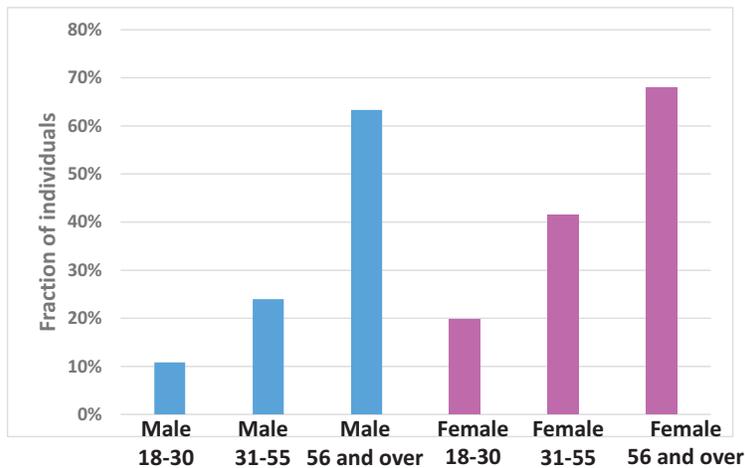}
\caption{Distribution of unsuccessful search for different age intervals and for each gender.}
\label{fig:unsuccessful}
\end{figure}

We call a search unsuccessful when (i) the result set $G$ is empty or (ii) the size of the anonymity set is above 100. In Figure~\ref{fig:unsuccessful}, we show the distribution of unsuccessful search for different age intervals and for each gender. As expected, the threat is stronger for males as it is easier to find the parents of a male compared to a married female (as married females mostly adopt the last names of their husband). Therefore, the inference algorithm returns a successful search more for males compared to females. Furthermore, algorithm returns unsuccessful searches mostly for older victims as the algorithm needs to trace one or two generations to find the result and such information is mostly unavailable for older individuals (e.g., due to deceased parents that do not exist in the database).

\begin{figure}[ht]
\centering
\includegraphics[scale=0.75]{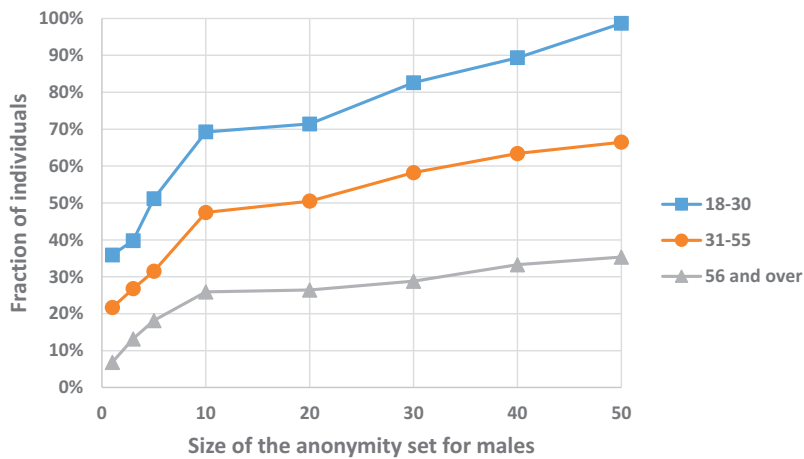}
\caption{Size of the anonymity set for males with different age ranges.}
\label{fig:male}
\end{figure}

\begin{figure}[ht]
\centering
\includegraphics[scale=0.75]{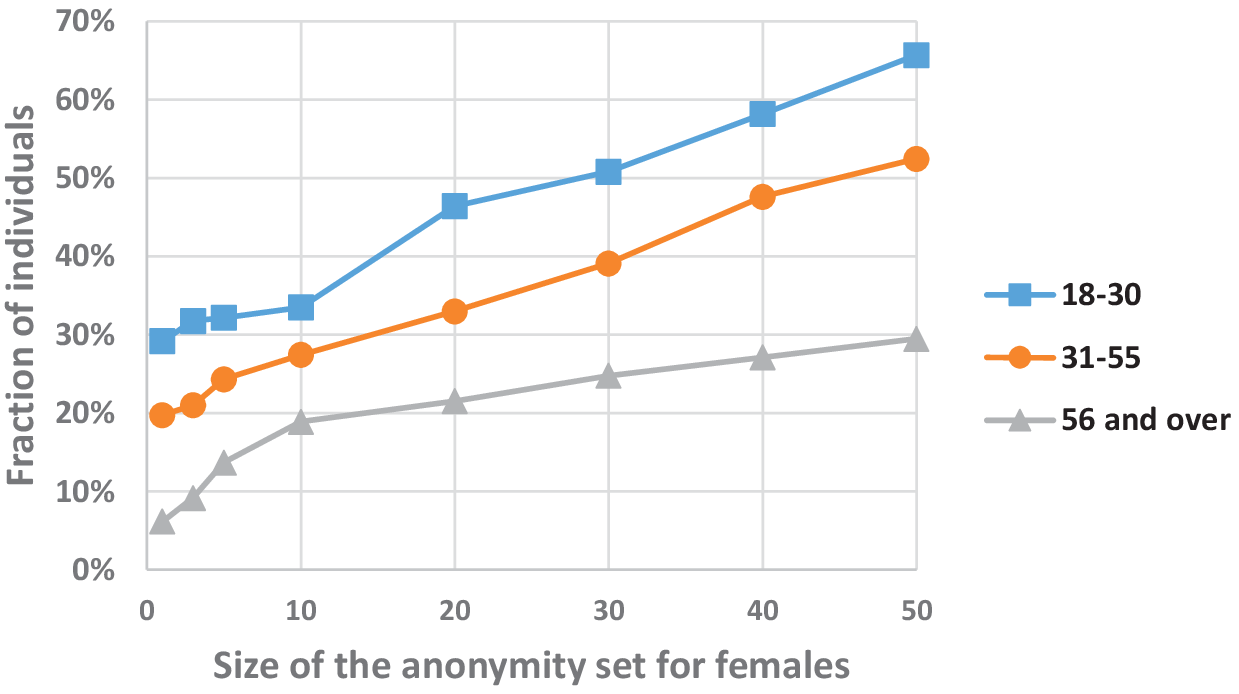}
\caption{Size of the anonymity set for females with different age ranges.}
\label{fig:female}
\end{figure}

In Figures~\ref{fig:male} and~\ref{fig:female}, we only focus on the successful searches and show the size of the obtained anonymity set for different age intervals and for each gender (numbers in these figures are also shown in Table~\ref{table:maiden}). In general, we observe that on the average, the size of the anonymity set is smaller for males and for younger victims. Also, we observe that for $69.25\%$ of the sampled individuals that are male and in the age range of $18-30$ the size of the anonymity set is below $10$.


Note that we did this study by only using the dataset and without using any other auxiliary information. Inference can be even stronger by using publicly available data on online social networks.\footnote{Married female users prefer declaring their maiden names on social networks (e.g., to find their friends from school).} We will also consider such auxiliary information in future work and will show how the size of the anonymity set changes with such auxiliary information.
\begin{table}[ht]
\centering
\includegraphics[scale=0.75]{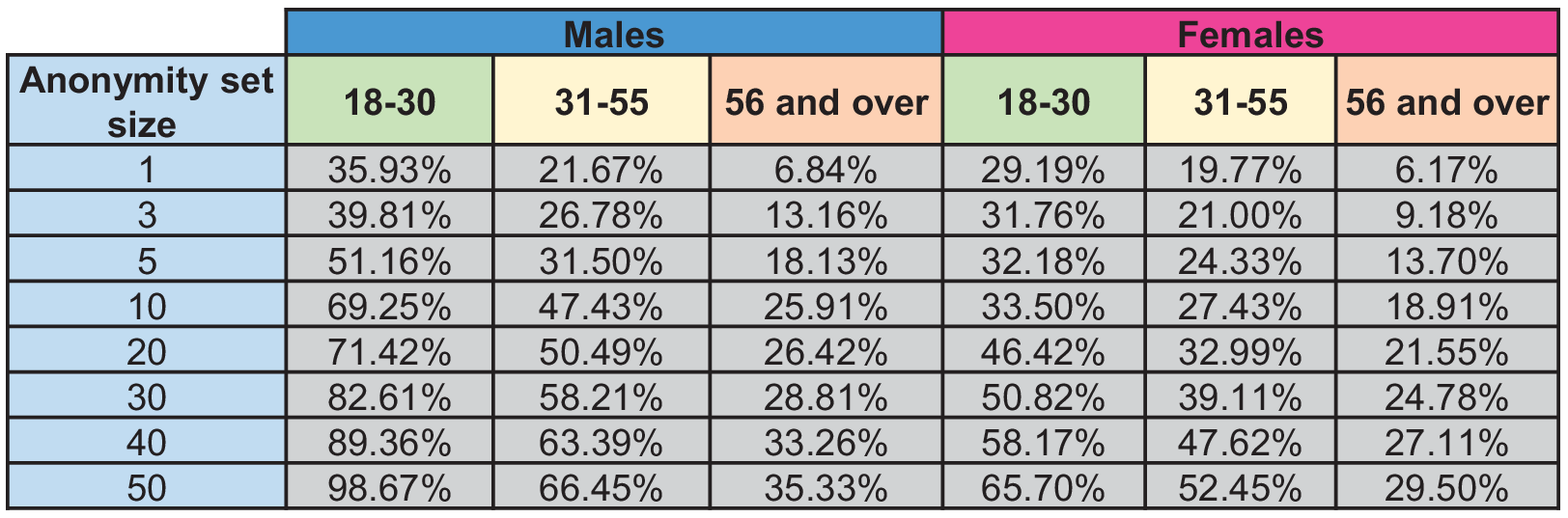}
\caption{Fractions of individuals in each anonymity set based on their demographics (i.e., age range and gender).}
\label{table:maiden}
\end{table}

\subsection{Inferring Landline Numbers}\label{sec:phone}

Turkish terror phone scams has been a major issue in the recent years and have attracted global attention~\cite{phone_scam}. Typically, scammers call people, claim they are police officers and tell them that their (or a relative's) name has showed up in a terror investigation. They also say that they know the victim is innocent but need their help and they are supposed to deposit a certain amount of money into a specified bank account to clear their names. They even play police radio sounds in the background to convince people that they are calling from the police station. Even a famous professor in Turkey has been a victim of this scam as well~\cite{phone_scam}.

Landline phone numbers and addresses of individuals were used to be released in thick hard-copy books called ``Golden Directory'' in the 90s in Turkey. Even though slightly out-dated, such directories are still publicly available. Today, the main provider of landline phones in Turkey, Turk Telekom,  provides an online lookup service named ``White Pages''~\cite{telekom}. This is a captcha protected service in which a user can query a last name in conjunction with an area code. Up to 20 individuals along with their landline numbers and addresses (number, neighborhood, district and city) are listed. To see 20 more individuals, user is requested to pass another captcha test. If there are too many matching last names, users are requested to enter additional information to narrow the query result (e.g., first name). Thus, scammers are able to address individuals by their names on the phone, which gives them credibility to a certain degree that they are who they claim to be. So far, the scammers were not able to provide any other credentials such as national identifier, date of birth, or birth place. However, this has potential to change with the recent leak. Now, a scammer has access to various extra private information about potential targets. Attackers can link the landline number and address of an individual (which are public) to the leaked dataset. The only obstacle for an attacker is the possibly large number of matching records in the phone directory due to namesakes.

In order to determine uniquely identifiable phone numbers given a name, we have randomly retrieved 983 phone numbers from the White Pages. Then, we checked the number of unique entries found in the leaked dataset for each landline number to determine the percentage of uniquely identifiable individuals. Out of the retrieved 983 records, we could find 702 of them in the leaked dataset.\footnote{281 entries could not be found possibly due to change of address. White Pages includes the recent addresses of individuals where data in the leaked dataset is from 2009.} Out of these 702 records, we observed that 60.11\% of the individuals are uniquely identifiable. Also, for 79.49\% of the individuals, we observed that the size of the anonymity set is smaller than or equal to 2 and for 85.47\% of the individuals, the size of the anonymity set is smaller than equal to 3. These results mean that a scammer can randomly pick an individual from the public phone directory, call him and in addition to addressing him with his name, the scammer can provide personal information such as national identifier, date of birth, and birth place. Even if the scammer is unsuccessful at the first attempt for the people in 2-anonymity sets, he will be able to hang-up the phone and call the next victim. Moreover, as shown in Section~\ref{sec:maiden}, it is possible to (i) find relatives of individuals in the leaked dataset, and (ii) with high probability, determine his mother's maiden name. Thus, a scammer can enumerate the relatives of the victim and can even tell their mother's maiden name. Given that scammers were very successful at convincing people and stealing their money with information limited to only first and last name, current situation poses a huge security risk.

\section{Discussion and Future Work}\label{sec:discussion}

The recent Turkish citizen dataset leak has major privacy risks as the identities of around 50 million individuals can now be associated with their national identifier numbers, date of births, and birth places. We show in this paper that with simple automated processing of the data, we can infer more information about the individuals, and the situation is even worse than currently anticipated. Strict measures must be taken to ensure the privacy and security of Turkish citizens because (i) the authentication systems of various institutions such as banks or government agencies are currently vulnerable, and (ii) very detailed information about each individual is now available for the scammers.

Mother's maiden name is a convenient authentication mechanism. It is usually not shared with other people because it is out of context and only known by the individual. It is also easy to remember. However, it is also known by close relatives and this means it is still not a personal secret. This was a problem during the pre-internet era. In today's world where so many people use social media as a means of communication, it is very easy to track someone's relatives and access their mothers' maiden names. For mothers who retain their maiden name, this is even as simple as learning the target's mother's full name. This makes mother's maiden name even more accessible.

Having that said, in Turkey (and also in the World) mother's maiden name is still commonly used as an authentication scheme. The situation is now much worse for Turkish citizens due to the leak. We showed that it is possible to pinpoint the maiden name of the mother's of around 30\% of the individuals (in the age range of 18-30, regardless of the gender). We were able to do this in an automated way, and only by using the leaked dataset itself. An attacker can still find out the mother's maiden name of other targets by linking more external information to this leaked dataset such as Google search results or social media accounts.

Therefore, we suggest the institutions (especially the financial institutions), which use mother's maiden name as an authentication scheme, to basically stop using it. One alternative is to use 2-factor authentication for telephone banking. 2-factor authentication is widely used to obtain temporary passwords for internet banking. The idea is to use an independent communication channel that is known to be private to the user (e.g., a smart phone running with a specific simcard). For telephone banking, on the other hand, mother's maiden name is preferred, as it is easier to verify verbally. Given today all individuals own at least one cell phone, 2-factor authentication is reasonable to use for telephone banking and it should replace mother's maiden name as the security mechanism.

Another alternative can be the Verint Identity Authentication~\cite{verint}. Verint Identity Authentication and Verint Fraud Detection work by instantly comparing the caller's voice to the voice prints of an extensive database of known fraud perpetrators. This comparison is done silently in the background of a call. The solution combines a new generation of passive voice biometrics with unique predictive analysis that can help accurately detect fraudsters and authenticate customers without caller interruption. Voice recognition technology offers a powerful combination of reduced customer effort and fraud deterrence. We expect this adoption will rapidly increase, not just among financial institutions but also among insurance, e-commerce, utilities, and other security-sensitive businesses.

As stated earlier, scammers take advantage of naive citizens by convincing them that they are involved in a legal investigation regarding terrorists and they need to deposit money into a bank account to clear their names. The online phone directory provided by Turk Telekom enables scammers to address individuals by name and a generalized address only. We showed that by automatically linking the leaked dataset to a subset of Turk Telekom phone directory, we are able uniquely identify more than 60\% of the individuals. Hence, now an attacker can address individuals by their name plus all the information about them in the leaked dataset. Moreover, they can infer their mothers' maiden names and also gather information about their close relatives. Scammers have been able to trick a significant number of people by just using their names (in some cases even without using their names). Now with all the information publicly available, we expect the number of events to rise as scammers will now have more credibility. We recommend the authorities (e.g., law enforcement and the media) to warn Turkish citizens by all means against this attack. Also, the online phone directory should be more difficult to query. For instance, a query with only a last name should be disabled.

In future work, we will further study the privacy implications of the leaked dataset. For instance, linking publicly available social network data can yield more information about individuals and their relatives. This will require obtaining social network data, pre-processing it, and linking individuals with public profiles in the social network data with the leaked dataset. We will also analyze the threat smartphone apps pose (along with the information in the leaked dataset). Users tend to approve any data access request by apps if they really need the functionality app provides. Attackers, using these apps as trojan horses, can have access to valuable personal information about the users. Attackers can even have access to the text messages of the user, which may contain 2-factor authentication passwords. One other aspect of our analysis will be the threat analysis for e-government accounts in relation to this data leak. E-government accounts only  use the national identifier number and a user defined password. Should the password be weak, with the current leak many e-government accounts could be compromised. Based on the results of our future analyses, we are going to propose new security measures to protect individuals from possible threats.

\section{Conclusion}\label{sec:conclusion}

In this work, we demonstrated that the Turkish citizen dataset leak poses a larger threat than currently anticipated. The fact that the mother's maiden name is not directly provided has been a relief for the Turkish community. We aimed to perform automated attacks that show without any manual intervention we can extract valuable personal information. We showed that even in this scenario, disclosed information is much more than just national identifier numbers. We also showed how to uniquely determine the landline numbers of the individuals in the leaked dataset and proposed basic countermeasures for the presented attacks. Finally, we studied the uniqueness of the Turkish citizens based on their demographics. We believe that the outcome of this study will guide the researchers that need estimates of the threat of disclosing anonymized datasets along with demographics.

\bibliographystyle{abbrv}
\bibliography{Infer2016_arxiv}
\end{document}